# Design Reflections on Transition to LLM-Aided Novel Visualizations

**Richard Brath**
Uncharted Software Inc., Toronto, Canada
rbrath@unchartedsoftware.com
0000-0001-6006-2092

## ABSTRACT

This study examines data visualization design evolution over 12½ years, reflecting on the impact of Large Language Models over the last 3¾ years. Using a longitudinal corpus of 55 visualizations from a single-subject design record, the study identifies how LLMs have aided design-space exploration: reducing coding effort, enabling new design opportunities, shock, excitement, accomplishments, and shifts to the design process.

## INTRODUCTION

Large Language Models (LLMs) have fundamentally changed human-computer interaction, especially in how we use text. While much research focuses on LLMs for text generation, their impact on the *design process* of specialized analytical tools—in this case, text-focused data visualizations—remains under-explored.

Text-focused visualization is a sub-domain of data visualization with unique challenges including an evolving design space of potential representations, and, encoding data which may be non-numeric. In this field, *novelty* is defined as the development of new approaches across the visualization pipeline of *analytics*, *encodings*, *layouts*, and *interaction* models that expand the design space beyond current paradigms [1].

This study is a retrospective reflection, reviewing a corpus of 55 novel text-centric visualizations created by the author over 12½-years. As the author transitioned from academia (completing a PhD in the subject) [2] to industry, the primary artifact for this study is not a collection of sporadic journal articles but derived from a consistent series of monthly blog posts [3]. It is a design record of design intent and execution transitioning from manually-coded visualizations to increasingly-LLM aided implementations (the most recent 3¾ years).

## BACKGROUND

The author is a practitioner of design and data visualization for 35 years [4]. I have undergrad degrees in architecture [5] and computer science [6]. I have used computers to aid design over the time, from command-line precursors to CAD, to coding in low-level C; Visual Basic; JavaScript; and now LLMs.

Early work designing data visualizations taught me that hand-drawn sketches of visualizations left too much uncertainty. Is the concept implementable? Do characteristics intrinsic in the data create something that looks quite different? Will the expected patterns appear? These uncertainties are only resolved with a proof-of-concept implementation using real data.

I am also skeptical of visualization theory; having seen it emerge and evolve. Does the theory really capture what's feasible? (e.g. [7]). What really is the design space for visualization, beyond theoretic conceptions?

## METHOD

Throughout my career, I have grappled with understanding design spaces. I have acquired knowledge by making: through the design and implementation of novel design concepts. To reflect on broader patterns over time I assemble and arrange the design results spatially i.e. creating a visualization (e.g. [8]). This externalization of many prior design artifacts facilitates recall of the projects and decreases working memory load [9]. Below is a timeline visualization of the 55 novel text-centric visualizations from the blog at tiny size (larger version in supplemental materials). I print these artifacts at scale, including physical manipulation such as folding and sticky notes to facilitate review.

## OBSERVATIONS

The following are focused on the transformation of the design process pre/post LLM.

**Lower effort from concept ideation, exploration through validation**. Pre-LLM, exploration of a novel viz concept typically became a light-weight implementation effort, in Javascript+D3.js or Python. LLMs minimize the technical burden of learning libraries and writing coding, allowing for faster design

# TIMELINE OF NOVEL TEXT VISUALIZATIONS BY AUTHOR

Based on blog posts between 2013-2026

2013: Sep, Oct, Nov, Dec
2014: Jan, Feb, Mar, Apr, May, Jun, Jul, Aug (1), Sep (2), Oct, Nov, Dec (3)

Sep 2013. PhD to define the design space of textual visualizations begins. Blog begins. Prior to this point, text has been utilized naïvely in data visualizations as discussed in typical computer science curriculum and related research—typically as labels applied after the quantitative visualization has been created. Many early blog posts are related to background research, including examples of prior textual visualizations going back hundreds to years to aid identification of the breadth of the design space:

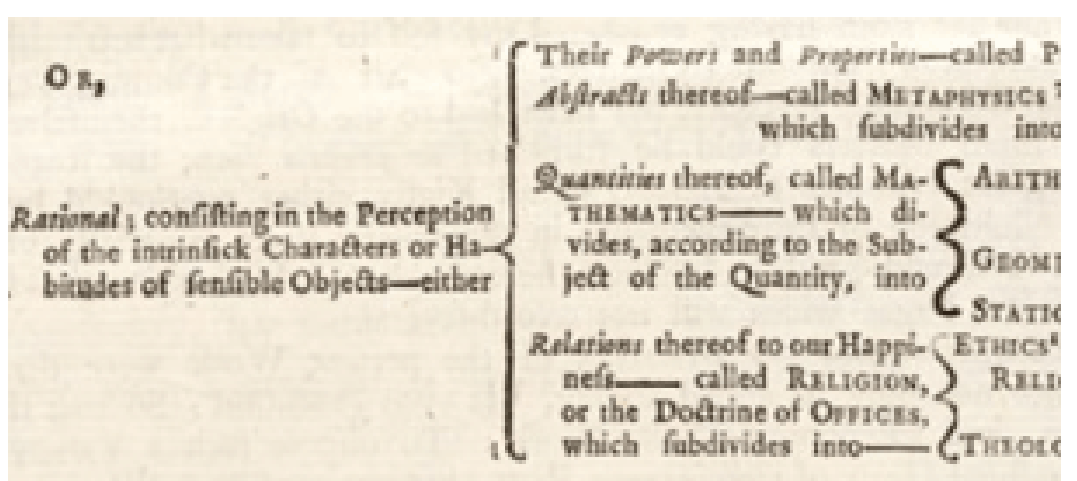

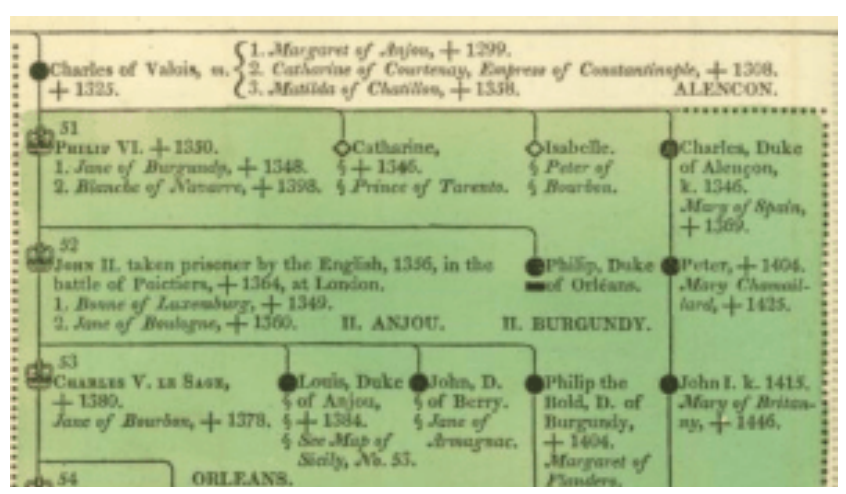

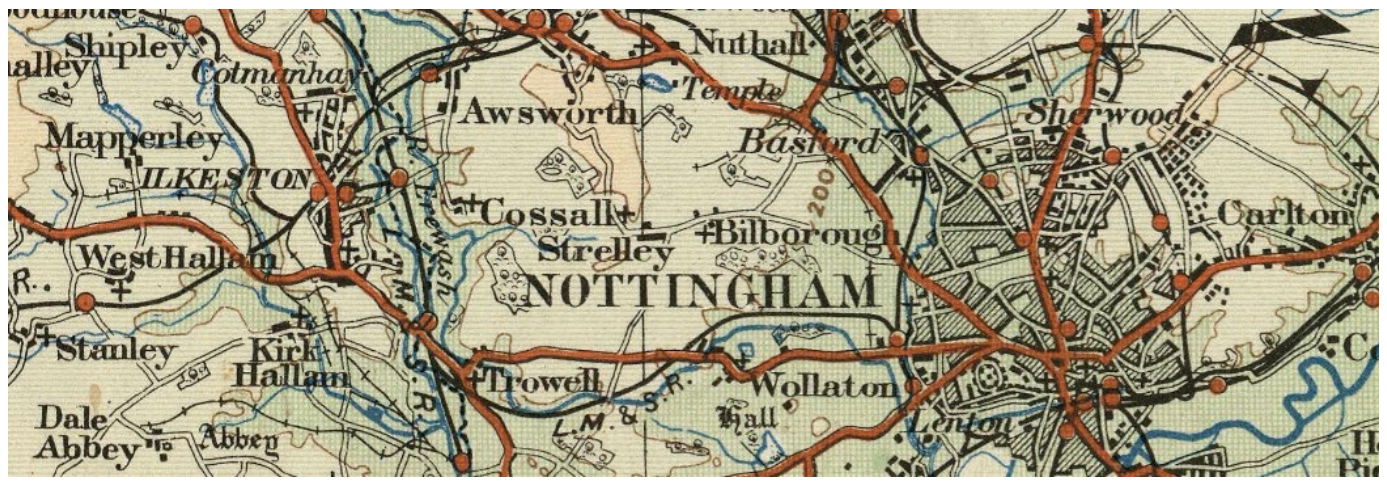

If the author's claims that the identified elements form a design space for the visualizations of textual data, then those design elements should be usable in novel designs. Hence, many blog posts are about new kinds of text-centric visualizations. The design investigation in most cases require software code to be manually written by author:

1. Equal area cartogram. (Aug 2014). *Opening salvo. Strong excitement.* Standard maps (choropleth, cartogram) don't show small entities. Adjusting geographic areas so all country codes are visible, small countries are be seen, and their values are comparable. →

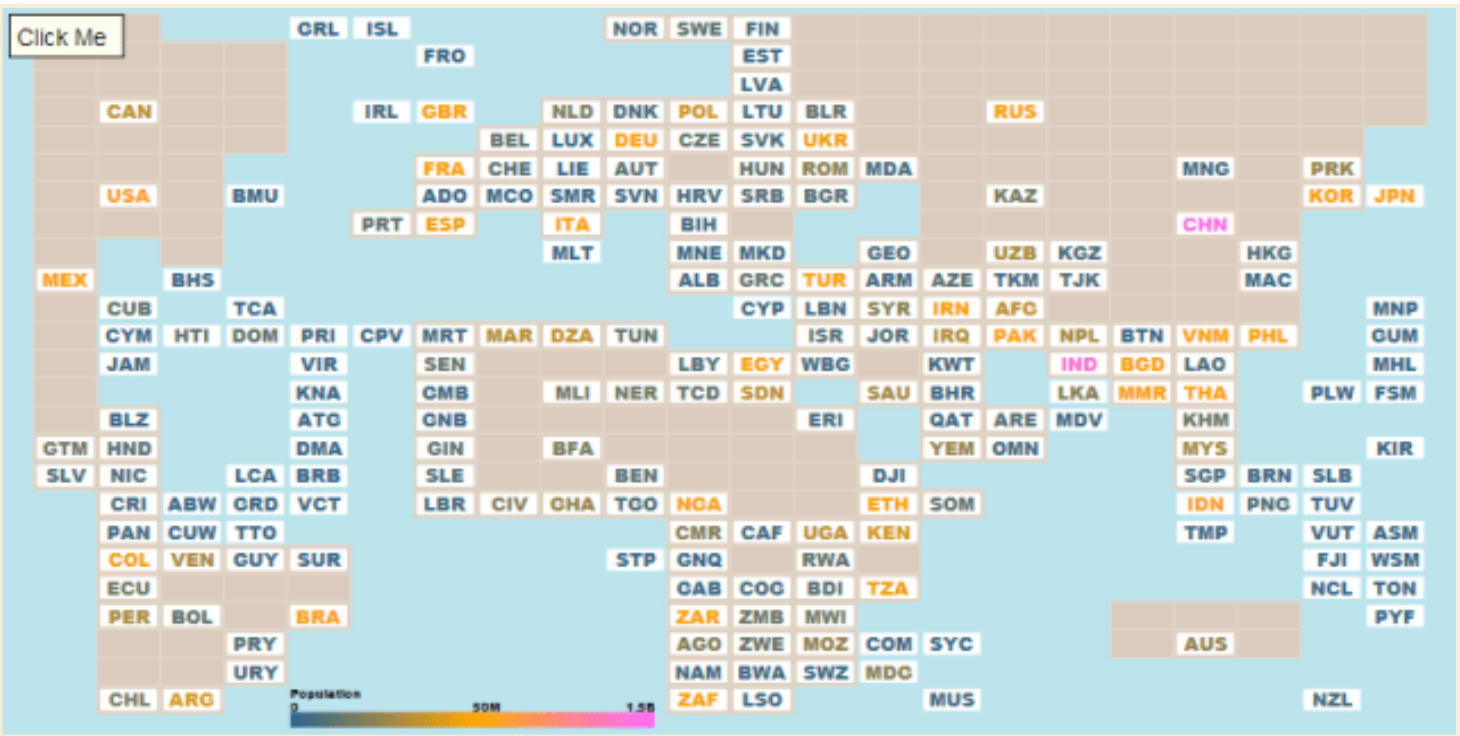

2. Positional encoding. (Sep 2014). Strange logic to push the envelope. A *line* of text has length, thus quantitative values can be marked and shown in relation to that length, as shown in this sentence. *How should this fit viz theory?* →

More[MEX] than[BRA] [IDN]5[ZAR]0[EGY]0 million women[ETH] in[BGD] the[NGA] world are[PAK] il[CHN]literate: the superscripts on this sentence indicate the number of illiterate women per country with sentence start indicating zero and sentence end indicating three hundred million illiterate [IND]women.

Data sources: UNESCO, World Bank, Wikipedia. MEX: Mexico. BRA: Brazil. IDN: Indonesia. ZAR: Democratic Republic of Congo. EGY: Egypt. BGD: Bangladesh. NGA: Nigeria. PAK: Pakistan. CHN: China. IND: India.
Author: Richard Brath, 2013

## Alice's Adventures in Wonderland

formatted for skimming using weight and italics

**Alice** was beginning *to* get very **tired** *of* sitting *by her* sister *on the* **bank,** *and of* having *nothing to* do. Once *or* **twice** *she* had **peeped** *into the* book *her* sister was reading, *but it* had *no* **pictures** *or* **conversations** *in it,* "*and* what *is the* use *of a* book," thought **Alice,** "*without* **pictures** *or* **conversations?"**

*So she* was **considering** *in her* own mind *(as* well *as she* could, *for the* day made *her* feel very **sleepy** *and* **stupid),** *whether the* pleasure *of* making *a* **daisy-chain** would be *worth the* trouble *of* getting up *and* **picking** *the* **daisies,** *when* suddenly *a* White **Rabbit** *with* **pink** eyes ran close *by her.*

There was *nothing* so very **remarkable** *in that, nor* did **Alice** *think it* so very much out *of the* way *to* hear *the* **Rabbit** say *to itself,* **"Oh dear! Oh dear!** *I* shall be too late!" *But when the* **Rabbit actually** took *a* watch out *of its* **waistcoat-pocket** *and* looked *at it and* then **hurried** *on,* **Alice** started *to her* feet, *for it* **flashed** *across her* mind *that she* had never *before* seen *a* **rabbit** *with* either *a* **waistcoat-pocket,** *or a* watch *to* take out *of it, and,* **burning** *with* **curiosity,** *she* ran *across the* field *after it and* was just *in* time *to* see *it* **pop** down *a* large **rabbit-hole,** *under the* **hedge.** *In* another moment, down went **Alice** *after it!*

*The* **rabbit-hole** went straight *on like a* **tunnel** *for* some **way** *and* then **dipped** suddenly down, so suddenly *that* **Alice** had not *a* moment *to* think *about* **stopping** *herself before she* found *herself* **falling** down what seemed *to* be *a* very deep well.

Either *the* well was very deep, *or she* fell very slowly, *for she* had *plenty of* time, *as she* went down, *to* look *about her.* First, *she* tried *to* make out what *she* was coming *to, but it* was too dark *to* see *anything;* then *she* looked *at the* **sides** *of the* well *and* **noticed** *that they* were filled *with* **cupboards** *and* **book-shelves;** here *and* there *she* saw **maps** *and* **pictures** hung *upon* **pegs.** *She* took down *a* **jar** *from* one *of the* **shelves** *as she* passed. *It* was **labeled "ORANGE MARMALADE,"** *but, to her* great **disappointment,** *it* was **empty;** *she* did not *like to* **drop** *the* **jar,** so **managed** *to* put *it into* one *of the* **cupboards** *as she* fell past *it.*

Down, down, down! Would *the* fall never come *to* an end? There ... *herself.* **"Dinah'll miss** *me* very much **to-night,** *I* should think!" ... **saucer** *of* milk *at* **tea-time. Dinah,** *my* dear, *I* wish *you* were dov... *when* suddenly, **thump! thump!** down *she* came *upon a* heap *of*...

| Font **weight** by word frequency: | | Font *italics* for: |
|---|---|---|
| light | top 100 | *articles,* |
| regular | 100-1000 | *conjunctions,* |
| **bold** | 1000-20000 | *prepositions,* |
| **black** | > 20000 | *pronouns,* |
| | | *infinitives* |

↑ 3. Skim formatting. (Dec 2014). Novel encoding of *inverse word frequency* to facilitate non-linear skimming of long texts, such as these initial paragraphs from *Alice in Wonderland*. The most uncommon words are most bolded so that they visually pop-out. *These last 3 designs aid reasoning that design space more than typographic attributes – also includes text scope (letter, word, line, etc).*

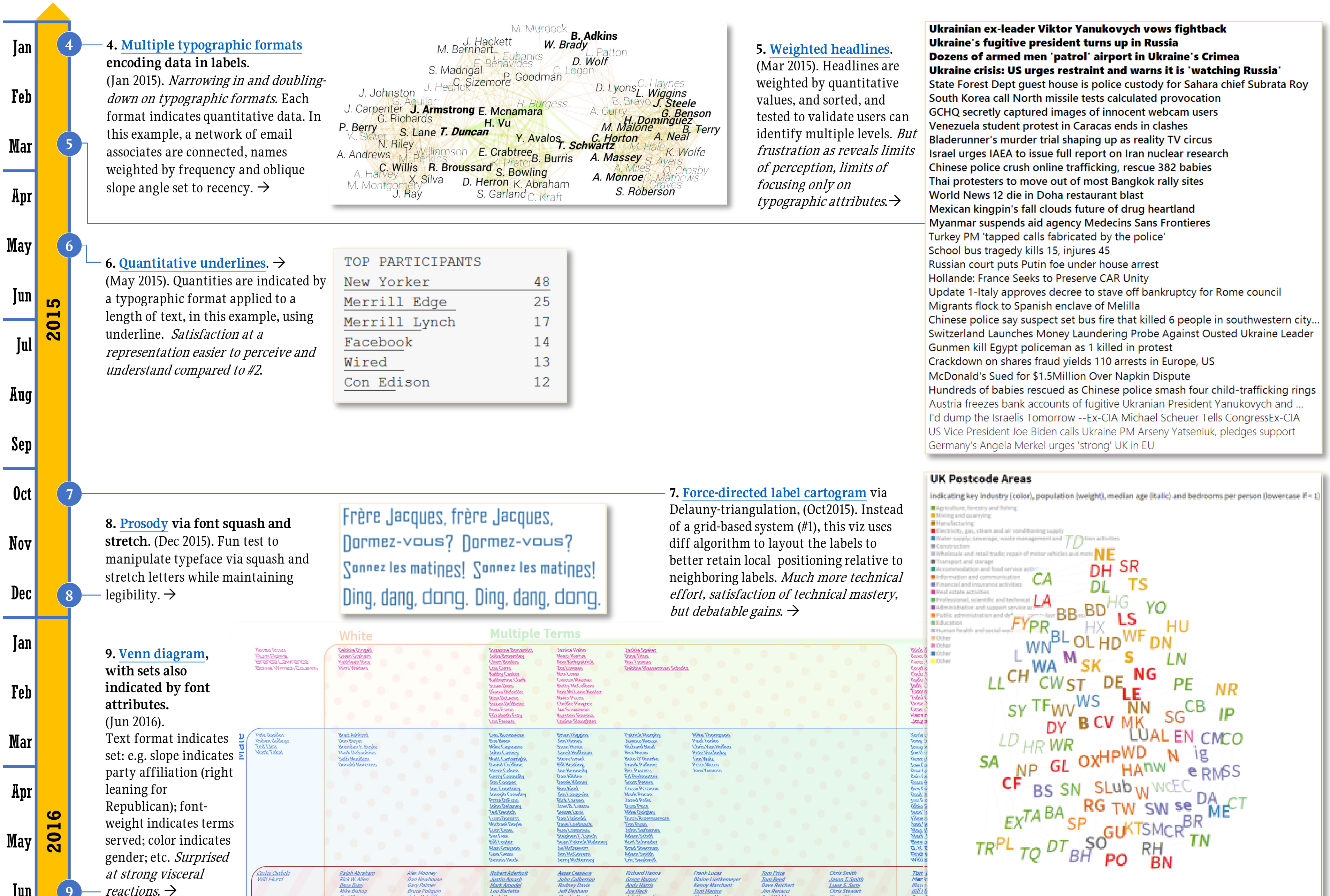
Jan
Feb
Mar
Apr
May
Jun
Jul
Aug
Sep
Oct
Nov
Dec
2015
Jan
Feb
Mar
Apr
May
Jun
2016
4
5
6
7
8
9
4. Multiple typographic formats encoding data in labels.
(Jan 2015). Narrowing in and doubling-down on typographic formats. Each format indicates quantitative data. In this example, a network of email associates are connected, names weighted by frequency and oblique slope angle set to recency. →
M. Murdock
B. Adkins
J. Hackett
W. Brady
M. Barnhart
L. Patton
D. Wolf
S. Madrigal
P. Goodman
C. Sizemore
J. Johnston
D. Lyons
L. Wiggins
J. Steele
J. Carpenter
J. Armstrong
E. Mcnamara
G. Richards
J. Benson
H. Dominguez
P. Berry
S. Lane
T. Duncan
H. Vu
M. Malone
B. Terry
Y. Avalos
C. Horton
A. Neal
N. Riley
T. Schwartz
A. Andrews
E. Crabtree
B. Burris
A. Massey
K. Wolfe
C. Willis
R. Broussard
S. Bowling
X. Silva
D. Herron
K. Abraham
A. Monroe
J. Ray
S. Garland
S. Roberson
5. Weighted headlines.
(Mar 2015). Headlines are weighted by quantitative values, and sorted, and tested to validate users can identify multiple levels. But frustration as reveals limits of perception, limits of focusing only on typographic attributes.→
Ukrainian ex-leader Viktor Yanukovych vows fightback
Ukraine's fugitive president turns up in Russia
Dozens of armed men 'patrol' airport in Ukraine's Crimea
Ukraine crisis: US urges restraint and warns it is 'watching Russia'
State Forest Dept guest house is police custody for Sahara chief Subrata Roy
South Korea call North missile tests calculated provocation
GCHQ secretly captured images of innocent webcam users
Venezuela student protest in Caracas ends in clashes
Bladerunner's murder trial shaping up as reality TV circus
Israel urges IAEA to issue full report on Iran nuclear research
Chinese police crush online trafficking, rescue 382 babies
Thai protesters to move out of most Bangkok rally sites
World News 12 die in Doha restaurant blast
Mexican kingpin's fall clouds future of drug heartland
Myanmar suspends aid agency Medecins Sans Frontieres
Turkey PM 'tapped calls fabricated by the police'
School bus tragedy kills 15, injures 45
Russian court puts Putin foe under house arrest
Hollande: France Seeks to Preserve CAR Unity
Update 1-Italy approves decree to stave off bankruptcy for Rome council
Migrants flock to Spanish enclave of Melilla
Chinese police say suspect set bus fire that killed 6 people in southwestern city...
Switzerland Launches Money Laundering Probe Against Ousted Ukraine Leader
Gunmen kill Egypt policeman as 1 killed in protest
Crackdown on shares fraud yields 110 arrests in Europe, US
McDonald's Sued for $1.5Million Over Napkin Dispute
Hundreds of babies rescued as Chinese police smash four child-trafficking rings
Austria freezes bank accounts of fugitive Ukranian President Yanukovych and ...
I'd dump the Israelis Tomorrow --Ex-CIA Michael Scheuer Tells CongressEx-CIA
US Vice President Joe Biden calls Ukraine PM Arseny Yatseniuk, pledges support
Germany's Angela Merkel urges 'strong' UK in EU
6. Quantitative underlines. →
(May 2015). Quantities are indicated by a typographic format applied to a length of text, in this example, using underline. Satisfaction at a representation easier to perceive and understand compared to #2.
TOP PARTICIPANTS
New Yorker 48
Merrill Edge 25
Merrill Lynch 17
Facebook 14
Wired 13
Con Edison 12
7. Force-directed label cartogram via Delauny-triangulation, (Oct2015). Instead of a grid-based system (#1), this viz uses diff algorithm to layout the labels to better retain local positioning relative to neighboring labels. Much more technical effort, satisfaction of technical mastery, but debatable gains. →
UK Postcode Areas
indicating key industry (color), population (weight), median age (italic) and bedrooms per person (lowercase if < 1)
8. Prosody via font squash and stretch. (Dec 2015). Fun test to manipulate typeface via squash and stretch letters while maintaining legibility. →
Frère Jacques, frère Jacques,
Dormez-vous? Dormez-vous?
Sonnez les matines! Sonnez les matines!
Ding, dang, dong. Ding, dang, dong.
9. Venn diagram, with sets also indicated by font attributes.
(Jun 2016).
Text format indicates set: e.g. slope indicates party affiliation (right leaning for Republican); font-weight indicates terms served; color indicates gender; etc. Surprised at strong visceral reactions. →
White
Multiple Terms

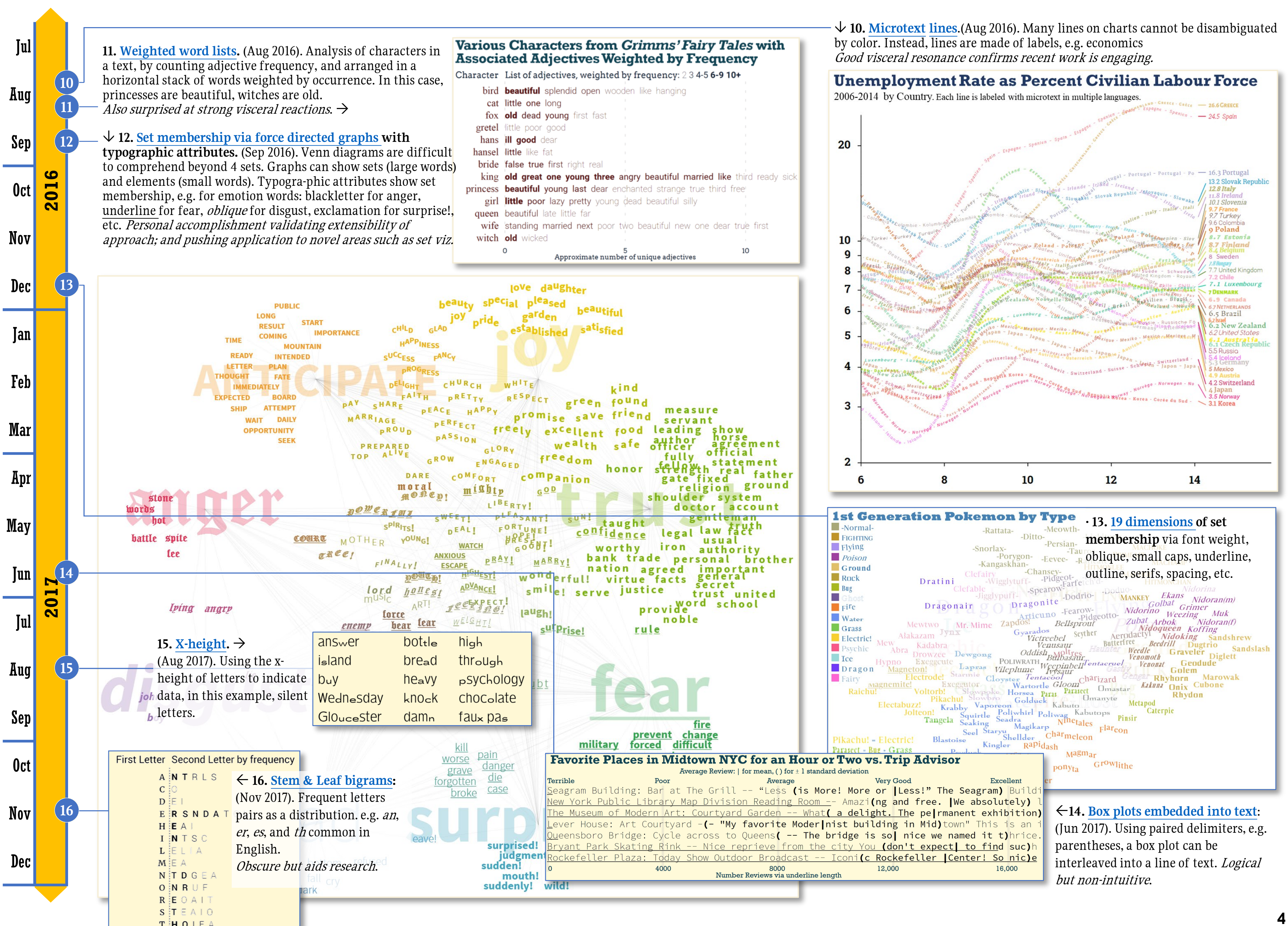

Jul
Aug
Sep
Oct
Nov
Dec
2016
Jan
Feb
Mar
Apr
May
Jun
Jul
Aug
Sep
Oct
Nov
Dec
2017
11. Weighted word lists. (Aug 2016). Analysis of characters in a text, by counting adjective frequency, and arranged in a horizontal stack of words weighted by occurrence. In this case, princesses are beautiful, witches are old.
Also surprised at strong visceral reactions. →
Various Characters from Grimms' Fairy Tales with Associated Adjectives Weighted by Frequency
Character List of adjectives, weighted by frequency: 2 3 4-5 6-9 10+
Approximate number of unique adjectives
↓ 10. Microtext lines.(Aug 2016). Many lines on charts cannot be disambiguated by color. Instead, lines are made of labels, e.g. economics
Good visceral resonance confirms recent work is engaging.
Unemployment Rate as Percent Civilian Labour Force
2006-2014 by Country. Each line is labeled with microtext in multiple languages.
↓ 12. Set membership via force directed graphs with typographic attributes. (Sep 2016). Venn diagrams are difficult to comprehend beyond 4 sets. Graphs can show sets (large words) and elements (small words). Typogra-phic attributes show set membership, e.g. for emotion words: blackletter for anger, underline for fear, oblique for disgust, exclamation for surprise!, etc. Personal accomplishment validating extensibility of approach; and pushing application to novel areas such as set viz.
1st Generation Pokemon by Type
• 13. 19 dimensions of set membership via font weight, oblique, small caps, underline, outline, serifs, spacing, etc.
15. X-height. →
(Aug 2017). Using the x-height of letters to indicate data, in this example, silent letters.
answer bottle high
island bread through
buy heavy psychology
Wednesday knock chocolate
Gloucester damn faux pas
First Letter Second Letter by frequency
← 16. Stem & Leaf bigrams:
(Nov 2017). Frequent letters pairs as a distribution. e.g. an, er, es, and th common in English.
Obscure but aids research.
Weight by frequency 0.5-0.75 0.75-1.0 1.0-1.5 1.5-2.0 >2.0%
Favorite Places in Midtown NYC for an Hour or Two vs. Trip Advisor
Average Review: | for mean, ( ) for ± 1 standard deviation
Number Reviews via underline length
←14. Box plots embedded into text:
(Jun 2017). Using paired delimiters, e.g. parentheses, a box plot can be interleaved into a line of text. Logical but non-intuitive.

Jan Feb Mar Apr May Jun Jul Aug Sep Oct Nov Dec — 2018

Jan Feb Mar Apr May Jun — 2019

Jul 2018. PhD successfully defended. Revisions completed the following month. A few items from the PhD continue to be posted shortly after this time.

→ **17.** Word stems. (Sep 2018). Commonality shown by spanning across all variants with taller (but narrower) text and/or rotated text. Visual structure of compound words vs. complex words apparent. *A quick design experiment inspired by #16, but more engaging with visceral reactions.*

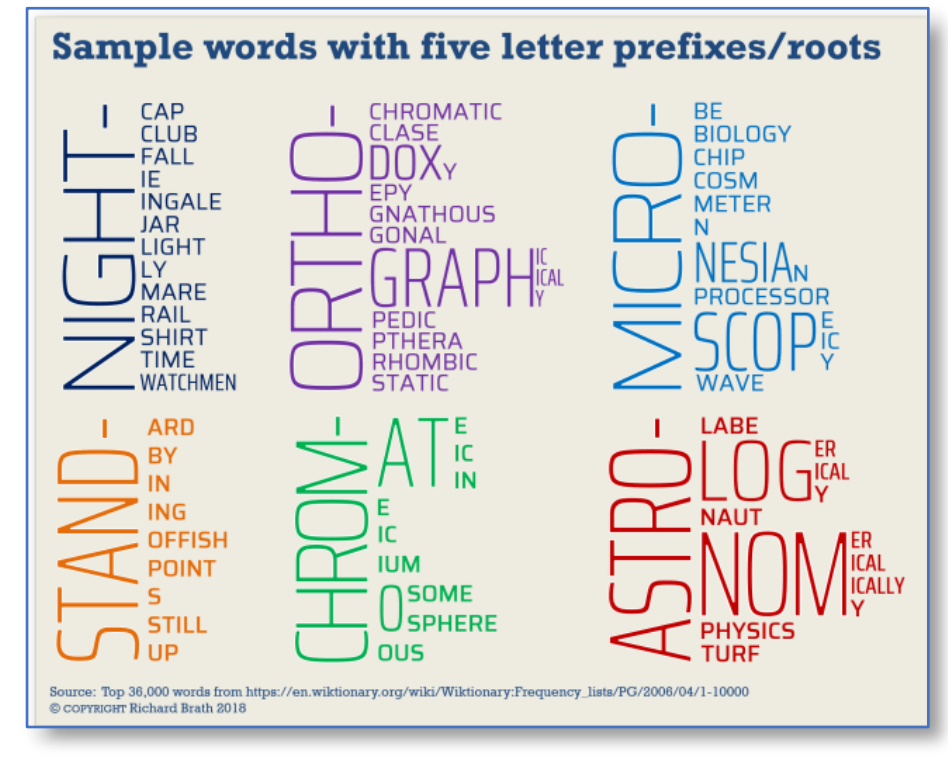

→ **18.** Letter-weight **represents data in that letter's geolocation.** Jan 2019. The breadth and orientation of text label indicates the extents of the category. The weight per each letter varies the data under that letter. *Another designed example with a non-intuitive result.*

Primary Occupation
Word indicates extents
Boldness indicates population
INDUSTRY SERVICES AGRICULTURE INDUSTRY AGRICULTURE SERVICES

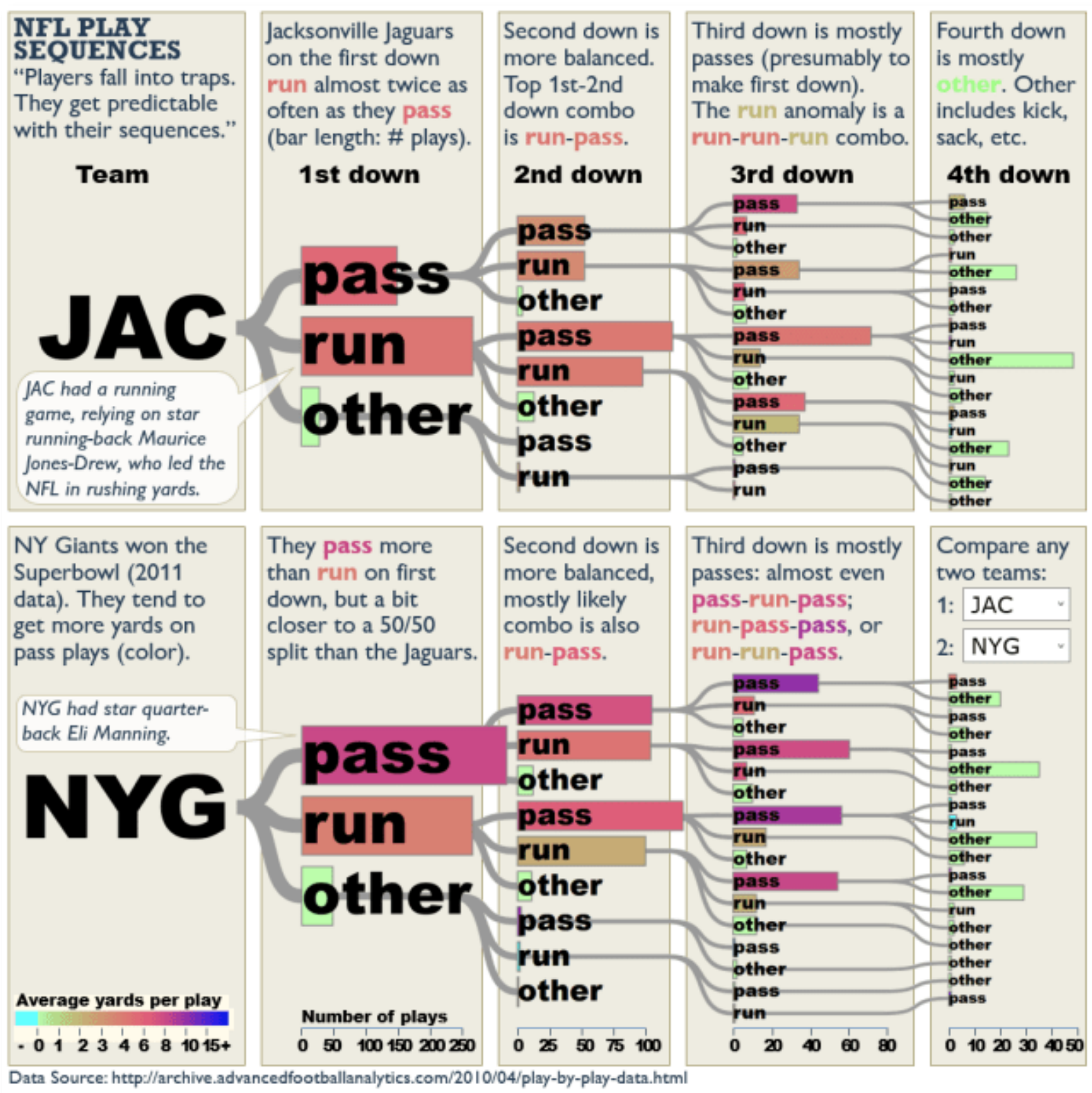

↑**19.** Textual data comics. Apr 2019. Data comics (B. Bach) borrow from traditional graphical narratives, combining data visualization, textual explanations, and sequential panels. In this case, *textual* visualization elements are used within each panel. In this example, each panel is a successive play by an NFL football team with mini-bar charts and explanation. *Uncertain of result: logical, engaging, but quite complex and narrative is tightly locked to panels.*

*Sparkwords.* (May 2019). Glyph-scale visualizations can be embedded inline with prose, popularized with *sparklines* (E. Tufte).

**20. Sparkwords via font attributes** embed data into words directly such as font weight or font color. These sparkwords can be embedded into prose. They may also cross-reference to other visualizations by common color, or other attributes. →

*This design late work, which occurred a year earlier, informed the designer and thesis to expand the design space into include literal text as a data type, for example, the encoded text was not meaningful outside its broader context.*

**21. Bars behind each letter.** (May 2019). More compactly, each letter conveys data, via a bar behind the letter, and reversing the fill color of the letter to maintain legibility at small sizes. In this example, the letters/bars convey baseball game scores and opponent teams. →

TOR TBR BAL BOS DET MIA TOR MIN LAA HOU CLE BOS OAK WSN KCR TEX LAA HOU BAL DET TOR NYM WSN TBR WSN SEA TBR PHL BOS ATL TOR BAL CLE NYM TBR KCR BAL BOS CHW TEX NYM TBR TOR MIA BAL CHW DET OAK SEA MIN TOR BOS BAL TBR BOS

In *Sémiologie graphique* (1967), Bertin illustrates examples with a dataset of occupations by department. Instead, observations can be visualized and explained in narrative. The top three departments by population (indicated by font weight: 34k-102, to 130, to 173, to 236, to 1517) are **Paris**, **Seine** and **Nord**. Top departments by percent of population in manufacturing (shown by proportion of red) are Belfort, **Moselle** and **Nord**. Top departments for agriculture (green) are Gers, Creuse and Lozere. Top services (blue) are **Paris**, Alpes Mmes, and **Bouches Du Rh**.

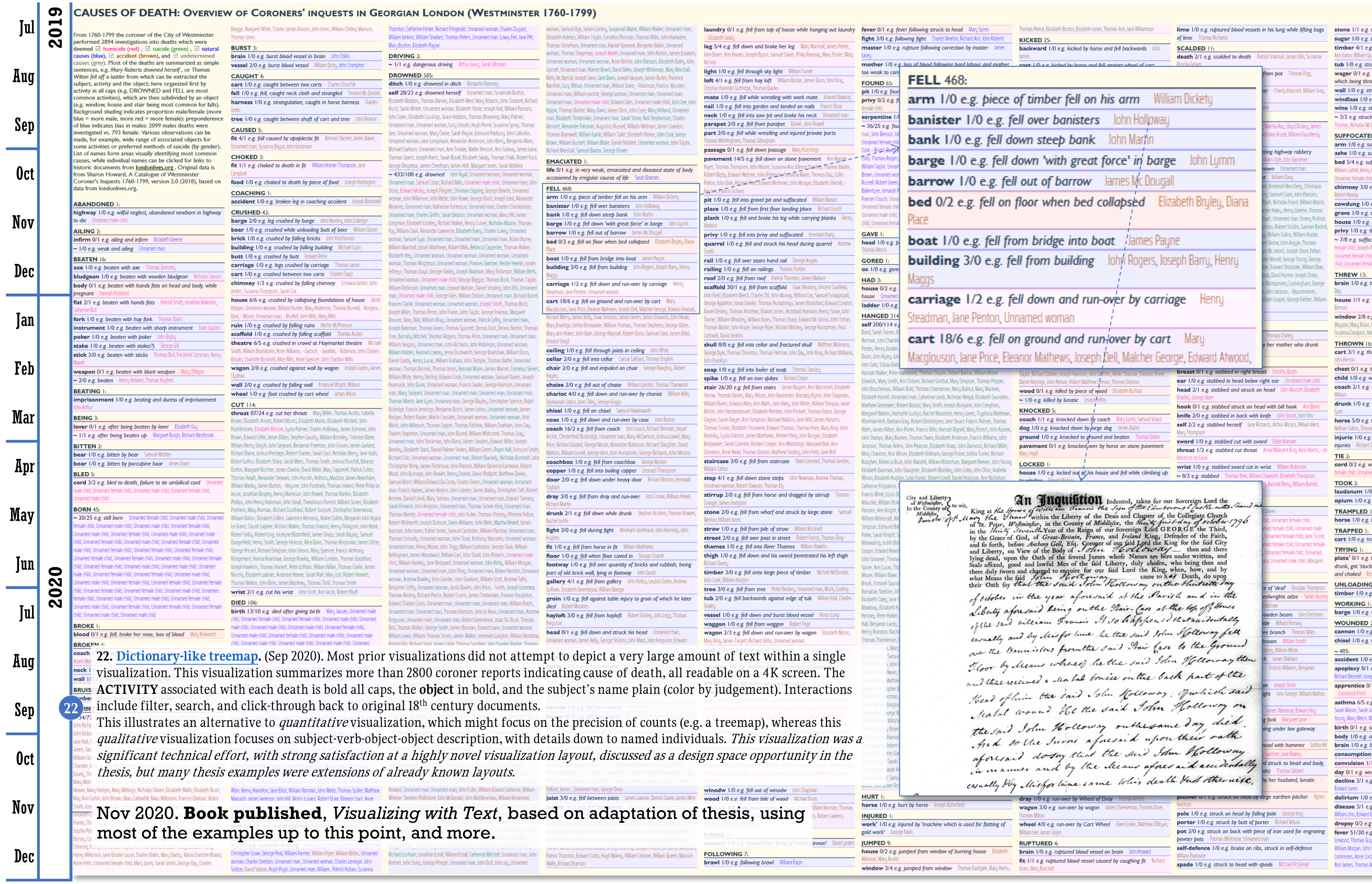

22. Dictionary-like treemap. (Sep 2020). Most prior visualizations did not attempt to depict a very large amount of text within a single visualization. This visualization summarizes more than 2800 coroner reports indicating cause of death, all readable on a 4K screen. The **ACTIVITY** associated with each death is bold all caps, the **object** in bold, and the subject's name plain (color by judgement). Interactions include filter, search, and click-through back to original 18th century documents.
This illustrates an alternative to *quantitative* visualization, which might focus on the precision of counts (e.g. a treemap), whereas this *qualitative* visualization focuses on subject-verb-object-object description, with details down to named individuals. *This visualization was a significant technical effort, with strong satisfaction at a highly novel visualization layout, discussed as a design space opportunity in the thesis, but many thesis examples were extensions of already known layouts.*

Nov 2020. **Book published**, *Visualizing with Text*, based on adaptation of thesis, using most of the examples up to this point, and more.

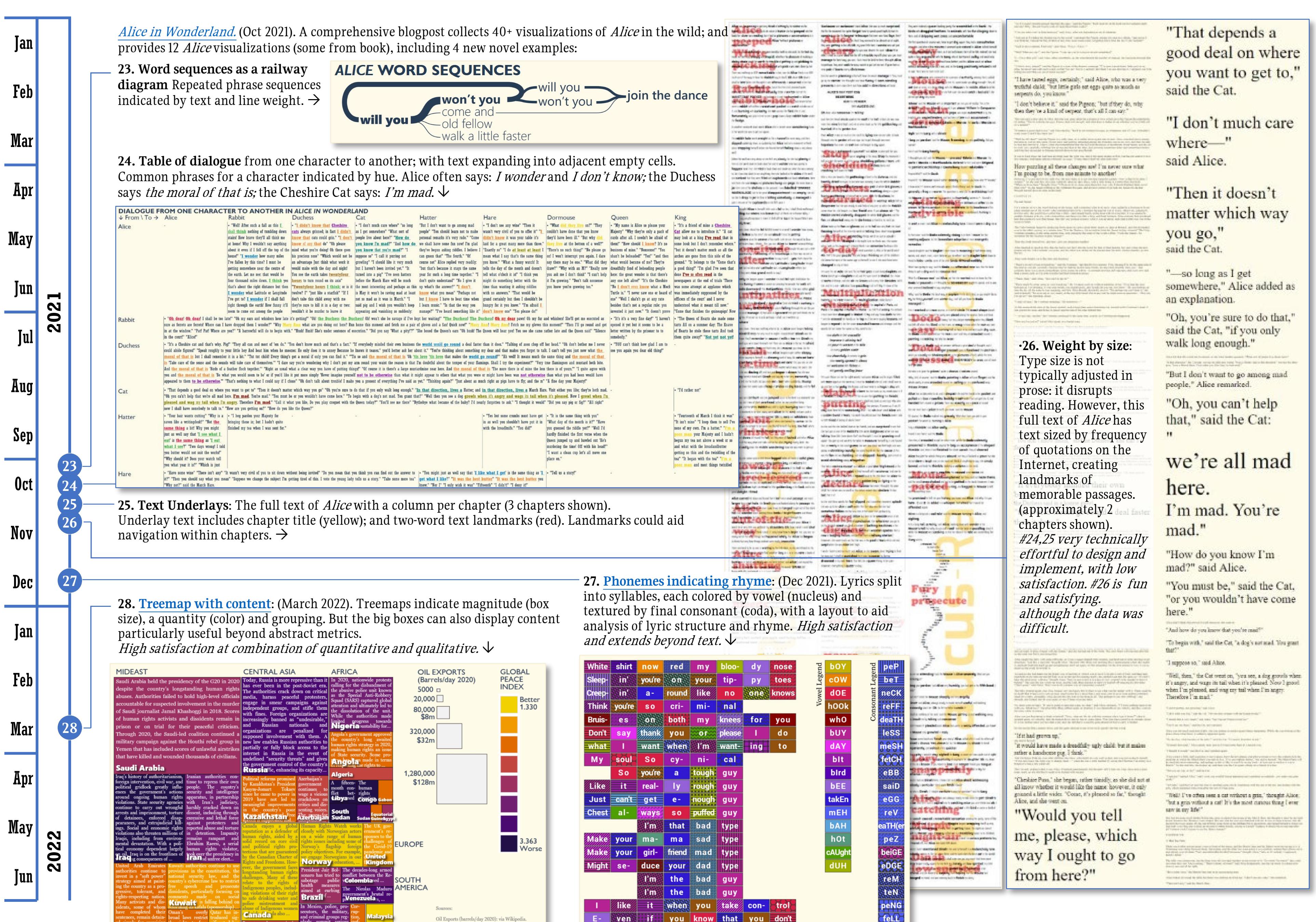
2021
Jan
Feb
Mar
Apr
May
Jun
Jul
Aug
Sep
Oct
Nov
Dec
2022
Jan
Feb
Mar
Apr
May
Jun
23
24
25
26
27
28
Alice in Wonderland. (Oct 2021). A comprehensive blogpost collects 40+ visualizations of Alice in the wild; and provides 12 Alice visualizations (some from book), including 4 new novel examples:
23. Word sequences as a railway diagram Repeated phrase sequences indicated by text and line weight. →
ALICE WORD SEQUENCES
will you
won't you
come and
old fellow
walk a little faster
will you
won't you
join the dance
24. Table of dialogue from one character to another; with text expanding into adjacent empty cells. Common phrases for a character indicated by hue. Alice often says: I wonder and I don't know; the Duchess says the moral of that is; the Cheshire Cat says: I'm mad. ↓
DIALOGUE FROM ONE CHARACTER TO ANOTHER IN ALICE IN WONDERLAND
↓ From \ To →
Alice
Rabbit
Duchess
Cat
Hatter
Hare
Dormouse
Queen
King
25. Text Underlays: The full text of Alice with a column per chapter (3 chapters shown). Underlay text includes chapter title (yellow); and two-word text landmarks (red). Landmarks could aid navigation within chapters. →
·26. Weight by size: Type size is not typically adjusted in prose: it disrupts reading. However, this full text of Alice has text sized by frequency of quotations on the Internet, creating landmarks of memorable passages. (approximately 2 chapters shown). #24,25 very technically effortful to design and implement, with low satisfaction. #26 is fun and satisfying, although the data was difficult.
"That depends a good deal on where you want to get to," said the Cat.
"I don't much care where—" said Alice.
"Then it doesn't matter which way you go," said the Cat.
"—so long as I get somewhere," Alice added as an explanation.
"Oh, you're sure to do that," said the Cat, "if you only walk long enough."
"Oh, you can't help that," said the Cat:
we're all mad here.
I'm mad. You're mad."
"How do you know I'm mad?" said Alice.
"You must be," said the Cat, "or you wouldn't have come here."
"Would you tell me, please, which way I ought to go from here?"
27. Phonemes indicating rhyme: (Dec 2021). Lyrics split into syllables, each colored by vowel (nucleus) and textured by final consonant (coda), with a layout to aid analysis of lyric structure and rhyme. High satisfaction and extends beyond text. ↓
Vowel Legend
Consonant Legend
28. Treemap with content: (March 2022). Treemaps indicate magnitude (box size), a quantity (color) and grouping. But the big boxes can also display content particularly useful beyond abstract metrics. High satisfaction at combination of quantitative and qualitative. ↓
MIDEAST
CENTRAL ASIA
AFRICA
OIL EXPORTS (Barrels/day 2020)
GLOBAL PEACE INDEX
Better 1.330
3.363 Worse
EUROPE
SOUTH AMERICA
NORTH AMERICA
LATIN AMERICA
PACIFIC
Saudi Arabia
Iraq
Iran
Russia
Nigeria
Angola
Algeria
Libya
Kazakhstan
Azerbaijan
Sudan
South Sudan
Congo
Norway
United Kingdom
Canada
Brazil
Colombia
Venezuela
Mexico
United States
Kuwait
United Arab Emirates
Oman
Qatar
Bahrain
Malaysia

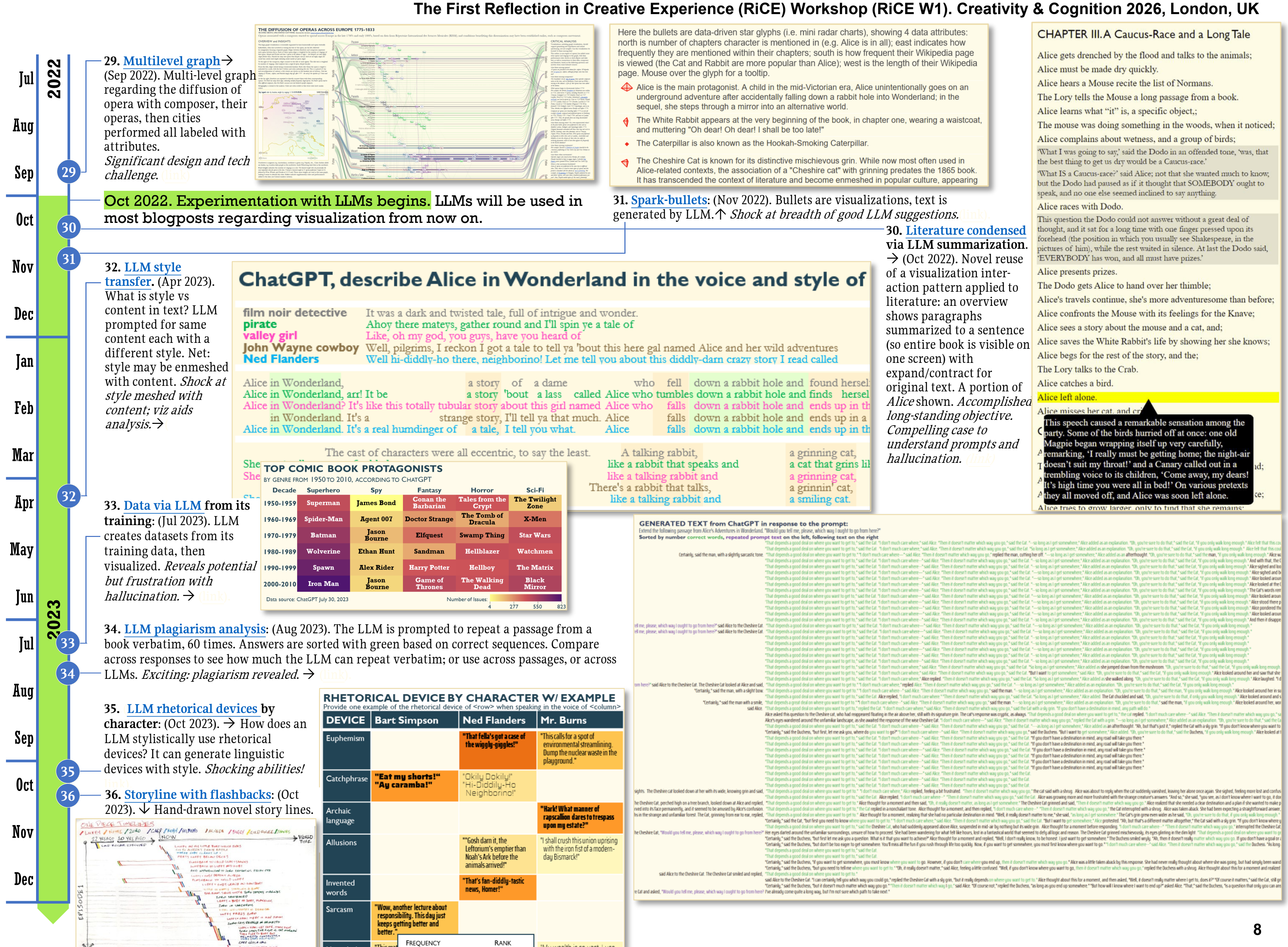

2022
Jul
Aug
Sep
Oct
Nov
Dec
2023
Jan
Feb
Mar
Apr
May
Jun
Jul
Aug
Sep
Oct
Nov
Dec
29. Multilevel graph→ (Sep 2022). Multi-level graph regarding the diffusion of opera with composer, their operas, then cities performed all labeled with attributes. Significant design and tech challenge. (link)
Oct 2022. Experimentation with LLMs begins. LLMs will be used in most blogposts regarding visualization from now on.
31. Spark-bullets: (Nov 2022). Bullets are visualizations, text is generated by LLM.↑ Shock at breadth of good LLM suggestions. (link)
30. Literature condensed via LLM summarization. → (Oct 2022). Novel reuse of a visualization interaction pattern applied to literature: an overview shows paragraphs summarized to a sentence (so entire book is visible on one screen) with expand/contract for original text. A portion of Alice shown. Accomplished long-standing objective. Compelling case to understand prompts and hallucination. (link)
32. LLM style transfer. (Apr 2023). What is style vs content in text? LLM prompted for same content each with a different style. Net: style may be enmeshed with content. Shock at style meshed with content; viz aids analysis.→
33. Data via LLM from its training: (Jul 2023). LLM creates datasets from its training data, then visualized. Reveals potential but frustration with hallucination. → (link)
34. LLM plagiarism analysis: (Aug 2023). The LLM is prompted to repeat a passage from a book verbatim, 60 times. Answers are sorted with green based on correct sequences. Compare across responses to see how much the LLM can repeat verbatim; or use across passages, or across LLMs. Exciting: plagiarism revealed. → (link)
35. LLM rhetorical devices by character: (Oct 2023). → How does an LLM stylistically use rhetorical devices? It can generate linguistic devices with style. Shocking abilities! (link)
36. Storyline with flashbacks: (Oct 2023). ↓ Hand-drawn novel story lines.
Here the bullets are data-driven star glyphs (i.e. mini radar charts), showing 4 data attributes: north is number of chapters character is mentioned in (e.g. Alice is in all); east indicates how frequently they are mentioned within their chapters; south is how frequent their Wikipedia page is viewed (the Cat and Rabbit are more popular than Alice); west is the length of their Wikipedia page. Mouse over the glyph for a tooltip.
Alice is the main protagonist. A child in the mid-Victorian era, Alice unintentionally goes on an underground adventure after accidentally falling down a rabbit hole into Wonderland; in the sequel, she steps through a mirror into an alternative world.
The White Rabbit appears at the very beginning of the book, in chapter one, wearing a waistcoat, and muttering "Oh dear! Oh dear! I shall be too late!"
The Caterpillar is also known as the Hookah-Smoking Caterpillar.
The Cheshire Cat is known for its distinctive mischievous grin. While now most often used in Alice-related contexts, the association of a "Cheshire cat" with grinning predates the 1865 book. It has transcended the context of literature and become entrenched in popular culture, appearing
CHAPTER III. A Caucus-Race and a Long Tale
Alice gets drenched by the flood and talks to the animals;
Alice must be made dry quickly.
Alice hears a Mouse recite the list of Normans.
The Lory tells the Mouse a long passage from a book.
Alice learns what "it" is, a specific object,;
The mouse was doing something in the woods, when it noticed;
Alice complains about wetness, and a group of birds;
Alice races with Dodo.
Alice presents prizes.
The Dodo gets Alice to hand over her thimble;
Alice's travels continue, she's more adventuresome than before;
Alice confronts the Mouse with its feelings for the Knave;
Alice sees a story about the mouse and a cat, and;
Alice saves the White Rabbit's life by showing her she knows;
Alice begs for the rest of the story, and the;
The Lory talks to the Crab.
Alice catches a bird.
Alice left alone.
This speech caused a remarkable sensation among the party. Some of the birds hurried off at once: one old Magpie began wrapping itself up very carefully, remarking, 'I really must be getting home; the night-air doesn't suit my throat!' and a Canary called out in a trembling voice to its children, 'Come away, my dears! It's high time you were all in bed!' On various pretexts they all moved off, and Alice was soon left alone.
ChatGPT, describe Alice in Wonderland in the voice and style of
film noir detective
pirate
valley girl
John Wayne cowboy
Ned Flanders
It was a dark and twisted tale, full of intrigue and wonder.
Ahoy there mateys, gather round and I'll spin ye a tale of
Like, oh my god, you guys, have you heard of
Well, pilgrims, I reckon I got a tale to tell ya 'bout this here gal named Alice and her wild adventures
Well hi-diddly-ho there, neighborino! Let me tell you about this diddly-darn crazy story I read called
TOP COMIC BOOK PROTAGONISTS
BY GENRE FROM 1950 TO 2010, ACCORDING TO CHATGPT
Decade | Superhero | Spy | Fantasy | Horror | Sci-Fi
1950-1959 | Superman | James Bond | Conan the Barbarian | Tales from the Crypt | The Twilight Zone
1960-1969 | Spider-Man | Agent 007 | Doctor Strange | The Tomb of Dracula | X-Men
1970-1979 | Batman | Jason Bourne | Elfquest | Swamp Thing | Star Wars
1980-1989 | Wolverine | Ethan Hunt | Sandman | Hellblazer | Watchmen
1990-1999 | Spawn | Alex Rider | Harry Potter | Hellboy | The Matrix
2000-2010 | Iron Man | Jason Bourne | Game of Thrones | The Walking Dead | Black Mirror
Data source: ChatGPT July 30, 2023
Number of Issues:
GENERATED TEXT from ChatGPT in response to the prompt:
RHETORICAL DEVICE BY CHARACTER W/ EXAMPLE
DEVICE | Bart Simpson | Ned Flanders | Mr. Burns
Euphemism | | "That fella's got a case of the wiggly-giggles!" | "This calls for a spot of environmental streamlining. Dump the nuclear waste in the playground."
Catchphrase | "Eat my shorts!" "Ay caramba!" | "Okily Dokily!" "Hi-Diddily-Ho Neighborino!" |
Archaic language | | | "Hark! What manner of rapscallion dares to trespass upon my estate?"
Allusions | | "Gosh darn it, the Leftorium's emptier than Noah's Ark before the animals arrived!" | "I shall crush this union uprising with the iron fist of a modern-day Bismarck!"
Invented words | | "That's fan-diddly-tastic news, Homer!" |
Sarcasm | "Wow, another lecture about responsibility. This day just keeps getting better and better." | |
Hyperbole | | | "My wealth is so vast, I use $100 bills as kindling for my solid gold fireplace."
Imperious commands | | | "Release the hounds, post-haste!"
FREQUENCY
Almost always
Frequently
Often
Regularly/Habitually
Occasionally
RANK
1
2
3
4-6
7+

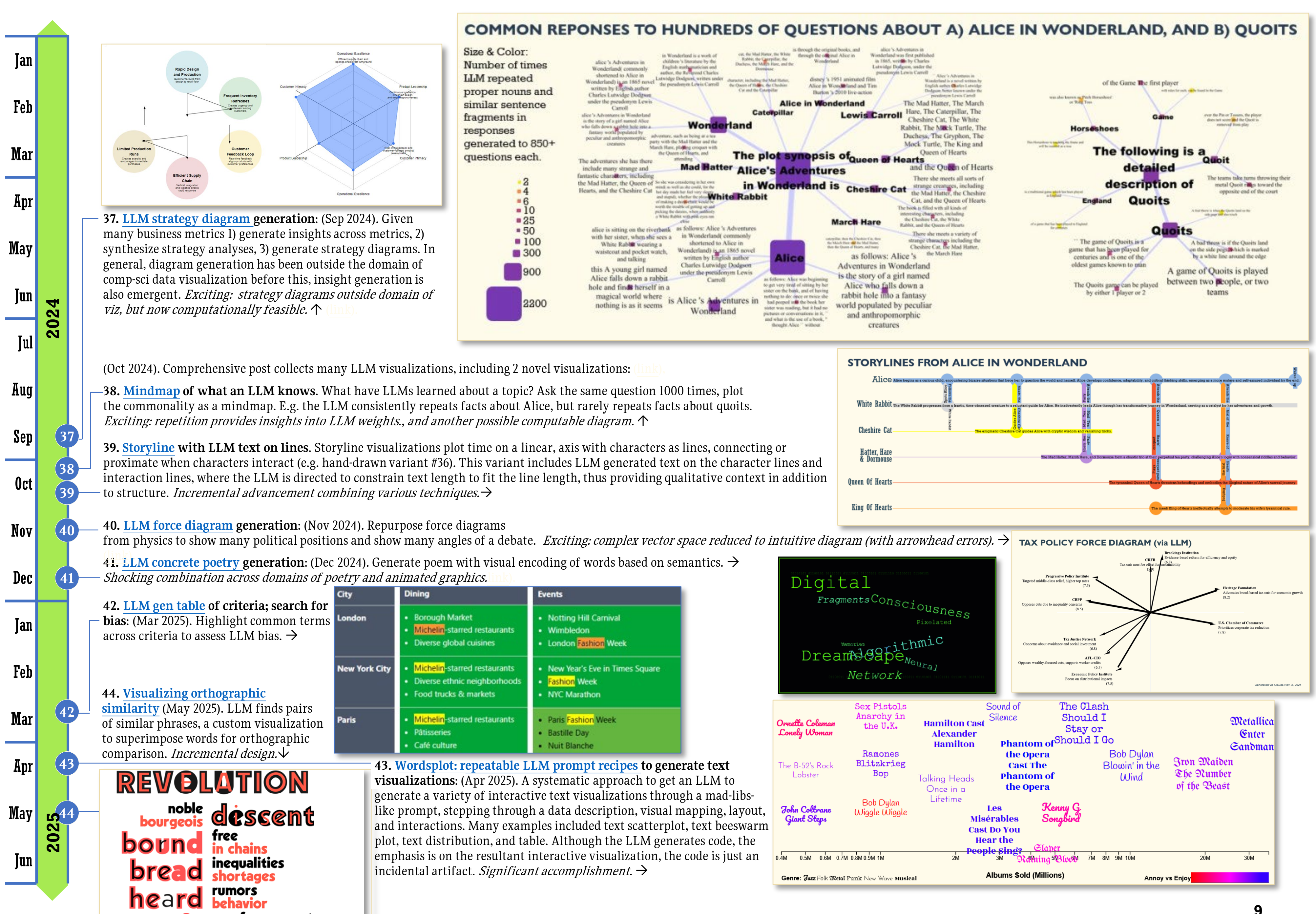

**37.** LLM strategy diagram **generation:** (Sep 2024). Given many business metrics 1) generate insights across metrics, 2) synthesize strategy analyses, 3) generate strategy diagrams. In general, diagram generation has been outside the domain of comp-sci data visualization before this, insight generation is also emergent. *Exciting: strategy diagrams outside domain of viz, but now computationally feasible.* ↑

(Oct 2024). Comprehensive post collects many LLM visualizations, including 2 novel visualizations:

**38.** Mindmap **of what an LLM knows**. What have LLMs learned about a topic? Ask the same question 1000 times, plot the commonality as a mindmap. E.g. the LLM consistently repeats facts about Alice, but rarely repeats facts about quoits. *Exciting: repetition provides insights into LLM weights., and another possible computable diagram.* ↑

**39.** Storyline **with LLM text on lines**. Storyline visualizations plot time on a linear, axis with characters as lines, connecting or proximate when characters interact (e.g. hand-drawn variant #36). This variant includes LLM generated text on the character lines and interaction lines, where the LLM is directed to constrain text length to fit the line length, thus providing qualitative context in addition to structure. *Incremental advancement combining various techniques.*→

**40.** LLM force diagram **generation:** (Nov 2024). Repurpose force diagrams from physics to show many political positions and show many angles of a debate. *Exciting: complex vector space reduced to intuitive diagram (with arrowhead errors).* →

**41.** LLM concrete poetry **generation:** (Dec 2024). Generate poem with visual encoding of words based on semantics. → *Shocking combination across domains of poetry and animated graphics.*

**42.** LLM gen table **of criteria; search for bias:** (Mar 2025). Highlight common terms across criteria to assess LLM bias. →

| City | Dining | Events |
|---|---|---|
| London | • Borough Market<br>• Michelin-starred restaurants<br>• Diverse global cuisines | • Notting Hill Carnival<br>• Wimbledon<br>• London Fashion Week |
| New York City | • Michelin-starred restaurants<br>• Diverse ethnic neighborhoods<br>• Food trucks & markets | • New Year's Eve in Times Square<br>• Fashion Week<br>• NYC Marathon |
| Paris | • Michelin-starred restaurants<br>• Pâtisseries<br>• Café culture | • Paris Fashion Week<br>• Bastille Day<br>• Nuit Blanche |

**44.** Visualizing orthographic similarity (May 2025). LLM finds pairs of similar phrases, a custom visualization to superimpose words for orthographic comparison. *Incremental design.*↓

**43.** Wordsplot: repeatable LLM prompt recipes **to generate text visualizations:** (Apr 2025). A systematic approach to get an LLM to generate a variety of interactive text visualizations through a mad-libs-like prompt, stepping through a data description, visual mapping, layout, and interactions. Many examples included text scatterplot, text beeswarm plot, text distribution, and table. Although the LLM generates code, the emphasis is on the resultant interactive visualization, the code is just an incidental artifact. *Significant accomplishment.* →

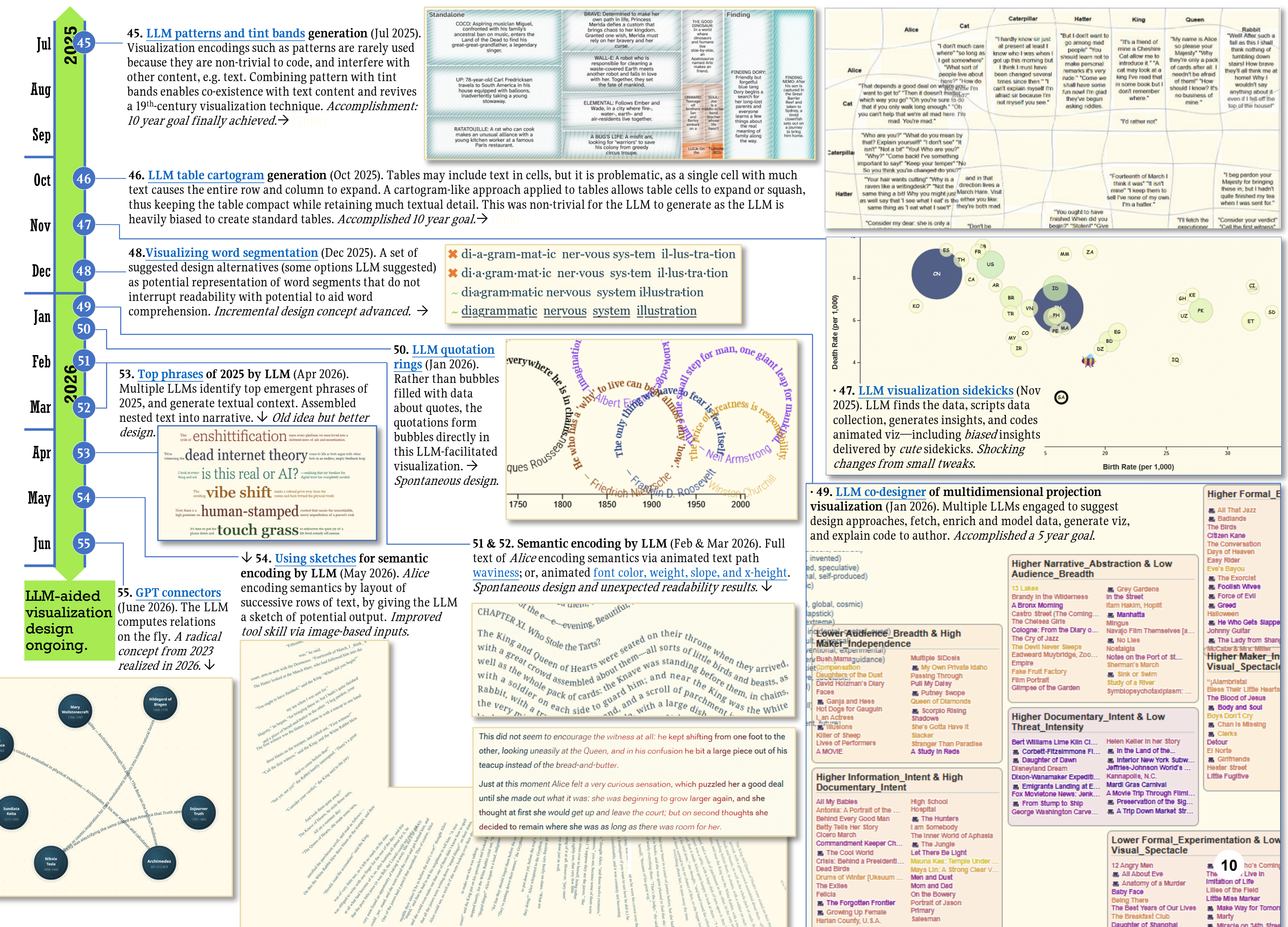
2025
Jul
Aug
Sep
Oct
Nov
Dec
2026
Jan
Feb
Mar
Apr
May
Jun
LLM-aided visualization design ongoing.
45. LLM patterns and tint bands generation (Jul 2025). Visualization encodings such as patterns are rarely used because they are non-trivial to code, and interfere with other content, e.g. text. Combining pattern with tint bands enables co-existence with text content and revives a 19th-century visualization technique. Accomplishment: 10 year goal finally achieved.→
46. LLM table cartogram generation (Oct 2025). Tables may include text in cells, but it is problematic, as a single cell with much text causes the entire row and column to expand. A cartogram-like approach applied to tables allows table cells to expand or squash, thus keeping the table compact while retaining much textual detail. This was non-trivial for the LLM to generate as the LLM is heavily biased to create standard tables. Accomplished 10 year goal.→
47. LLM visualization sidekicks (Nov 2025). LLM finds the data, scripts data collection, generates insights, and codes animated viz—including biased insights delivered by cute sidekicks. Shocking changes from small tweaks.
48. Visualizing word segmentation (Dec 2025). A set of suggested design alternatives (some options LLM suggested) as potential representation of word segments that do not interrupt readability with potential to aid word comprehension. Incremental design concept advanced. →
di-a-gram-mat-ic ner-vous sys-tem il-lus-tra-tion
di·a·gram·mat·ic ner·vous sys·tem il·lus·tra·tion
diagrammatic nervous system illustration
diagrammatic nervous system illustration
49. LLM co-designer of multidimensional projection visualization (Jan 2026). Multiple LLMs engaged to suggest design approaches, fetch, enrich and model data, generate viz, and explain code to author. Accomplished a 5 year goal.
50. LLM quotation rings (Jan 2026). Rather than bubbles filled with data about quotes, the quotations form bubbles directly in this LLM-facilitated visualization. → Spontaneous design.
51 & 52. Semantic encoding by LLM (Feb & Mar 2026). Full text of Alice encoding semantics via animated text path waviness; or, animated font color, weight, slope, and x-height. Spontaneous design and unexpected readability results. ↓
53. Top phrases of 2025 by LLM (Apr 2026). Multiple LLMs identify top emergent phrases of 2025, and generate textual context. Assembled nested text into narrative. ↓ Old idea but better design.
enshittification
dead internet theory
is this real or AI?
vibe shift
human-stamped
touch grass
↓ 54. Using sketches for semantic encoding by LLM (May 2026). Alice encoding semantics by layout of successive rows of text, by giving the LLM a sketch of potential output. Improved tool skill via image-based inputs.
55. GPT connectors (June 2026). The LLM computes relations on the fly. A radical concept from 2023 realized in 2026.↓
Mary Wollstonecraft
Hildegard of Bingen
Ada Lovelace
Sojourner Truth
Sundiata Keita
Nikola Tesla
Archimedes
1750
1800
1850
1900
1950
2000
Death Rate (per 1,000)
Birth Rate (per 1,000)
Higher Narrative_Abstraction & Low Audience_Breadth
Lower Audience_Breadth & High Maker_Independence
Higher Documentary_Intent & Low Threat_Intensity
Higher Information_Intent & High Documentary_Intent
Lower Formal_Experimentation & Low Visual_Spectacle
This did not seem to encourage the witness at all: he kept shifting from one foot to the other, looking uneasily at the Queen, and in his confusion he bit a large piece out of his teacup instead of the bread-and-butter.
Just at this moment Alice felt a very curious sensation, which puzzled her a good deal until she made out what it was: she was beginning to grow larger again, and she thought at first she would get up and leave the court; but on second thoughts she decided to remain where she was as long as there was room for her.
10

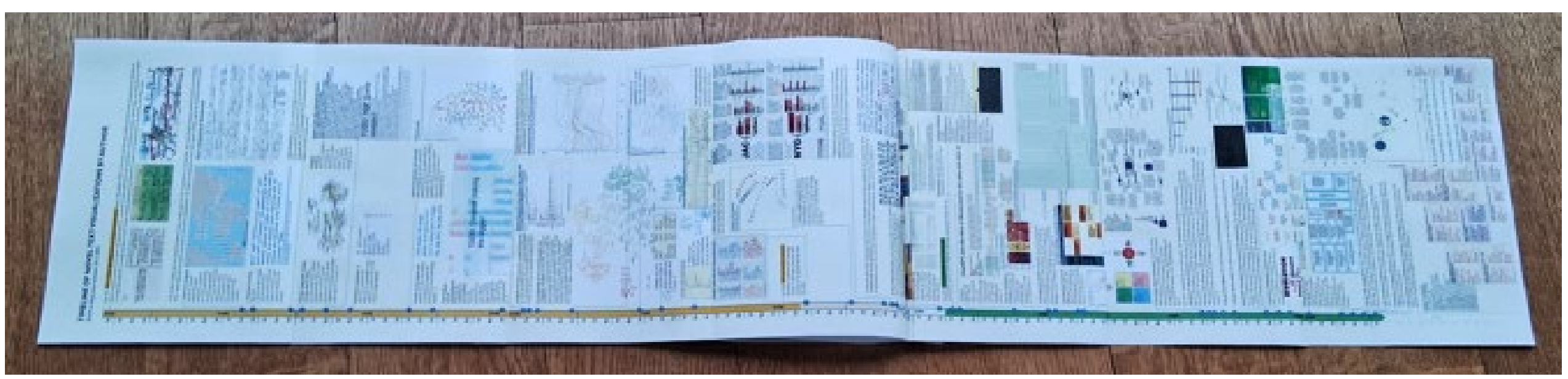

Draft of the timeline of text visualizations, printed and joined. The middle is folded to facilitate visual comparison across periods.

exploration in text analysis and visualization projects, a critical requirement when the author can only explore these design ideas on evenings and weekend. LLMs have ever increasing roles:
1) Natural language text processor (Oct 2022, #30).
2) Design collaborator (Nov 2022, #31)
3) System to be explained (Apr 2023, #32)
4) Data generator (Jul 2023, #33)
5) Extractor of insights from data (Sep 2024, #37)
6) Diagram generator (Sep 2024, #37)
7) Code generator (for novel designs) (Apr 2025, #43)
8) Explainer and annotator (Nov 2025, #47)
9) Interpreter of hand-drawn sketches (May 2026, #54)

Both pre/post, a project may span many weekends, due to frustration with tools (hand-written code, or LLM); and/or design ideation. Overall, the effort spent in design exploration as coding and/or prompting has reduced from days to hours, even considering LLM caveats such as hallucinations, imprecision, and other issues.

**Shocking capabilities** emerge with LLM use as a design tool, for what they can and can't do. In #30, the LLM suggested more design alternatives than I'd considered. The LLM generated creative text, such as metaphors (32), style-transfer (35), and concrete poetry (41) far beyond anticipated. In #47, it extracted highly different insights with a single adjective change. These findings can lead to serendipitous design opportunities (35,47). Also shocking: hallucinations (31, 33), simple data errors (47), and simple coding errors (54) which required due-diligence post-coding (or annotation). Ongoing familiarization with LLMs is required.

**Easy design exploration**. Pre-LLM, I was cautious pursuing some designs, as the core idea would require learning new technology needing much effort, before actual design exploration. With reduced level-of-effort afforded by LLMs, viz features explored included bendy tables (#46), patterned tint bands (#45), force diagrams (#40), animation (#47), condensed text to fit space (#39), rings of text (#50). Given the low risk and low cost to invoke an LLM with an off-the-cuff idea, spontaneous explorations are enabled (36,50,51,54).

**Excitement**. Working with an LLM-aided design tool has opened new areas for computational data viz, e.g. diagrams (37,38,40), close reading (44,53), and bendy tables (46). The latter was conceptualized 10 years earlier, but initial implementations were mediocre, followed by many unsuccessful attempts including LLMs (as LLMs are trained overwhelmingly on standard tables). Realizing long-term design goals brings a sense of accomplishment (31, 45, 49).

**LLMs have issues**: skipping data, hallucinations, graphical precision, etc. LLM-aided design users must know that models tend to produce bland, conventional, or average results, *unless explicitly directed by the designer with specificity* [10]. The designer is in-the-loop, not beside it.

**LLM context window saturation**. Due to accumulated incoherence as LLM conversations increase, design sessions must be planned and terse; with multiple new sessions. This changes the design process and requires various strategies, e.g. pre-prompt iteration, planning validation, arbitrary pauses.

**LLM-aided design is design**. A reviewer of a prior iteration of this paper asked: "How does the shift to *passive curatorial selection* feel?" suggesting the LLM use was not making per Cross [11]. To which I vehemently reply: coding with LLMs is coding, *not* passive curatorial selection. Crafting prompts, iteration, and code review is more akin to shifting from C to Visual Basic, or VB to Javascript + libraries. In none of the prior shifts did I feel I had reduced my role as designer nor maker, nor do I feel that now.

**LLM reflection**. All blog posts were sent to an LLM prompted for thematic differences post-LLM. Without specificity in the request, the LLM provided mundane analysis, such as earlier periods focused more on maps and typographic attributes. It further suggested that LLM usage could lead to generative homogenization – a risk also identified by the author above.

**Meta-reflection**. Reflection during PhD was highly focused to advancing the PhD, thus using design instantiations to revise and reframe the PhD. Post PhD and LLM periods explore the design space, pushing new opportunities afforded by lower effort LLM design, which in turn has exposed a potential new cycle of a broader reflection and reframing of a larger design space.

## CONCLUSIONS

LLMs have significantly changed the author's design process. LLMs aid my visualization design creativity including:

- **Code** is often required in data visualization, to validate and adapt design ideas to work with real data. Slow precise coding can be traded-off for fast iterative LLM-in-the-loop.
- **Subtasks** can be delegated to LLMs, such as finding data, data scraping, data processing, visual transformations, and so on.

- **Collaboration** with LLMs can aid design ideation, ranging from iterative exploration of loose design formulations, to a breadth of suggestions to break through the design equivalent of writers-block.
- **Whim and serendipity** are important in design [12]. LLMs aid rapid exploration of ideas that might otherwise be skipped or dismissed.

Overall, LLMs can facilitate design exploration and enable new classes of designs. This, in turn, will facilitate new uses, new applications, and new categories of software.

## ACKNOWLEDGEMENTS

The author's pictorial timeline, is open source, licensed under CC BY SA 4.0.